\documentclass[preprintnumbers, prd, twocolumn, showpacs, floatfix, preprintnumbers, 
letterpaper, 
superscriptaddress,nofootinbib]{revtex4-1}
\pdfoutput=1
\usepackage{graphicx,epsfig}
\usepackage{amsmath}
\usepackage {amssymb}
\usepackage {longtable}
\usepackage{multirow}

\usepackage{dcolumn}
\usepackage{bm}
\usepackage{amsfonts}
\usepackage{subfigure}
\usepackage{color}

\usepackage{relsize}
\usepackage{relsize}

\newcommand{\be}{\begin{equation}}
\newcommand{\ee}{\end{equation}}
\newcommand{\bea}{\begin{eqnarray}}
\newcommand{\eea}{\end{eqnarray}}

\def\DerN#1{\frac{d #1}{d N}}
\def\rf#1{(\ref{#1})}

\begin{document}


\title{Dynamical system analysis for a nonminimal torsion-matter coupled gravity}

\author{Sante Carloni}
\email{sante.carloni@tecnico.ulisboa.pt}
\affiliation{Centro Multidisciplinar de Astrofisica - CENTRA,
Instituto Superior Tecnico - IST,
Universidade de Lisboa - UL,
Avenida Rovisco Pais 1, 1049-001, Portugal}

\author{Francisco S. N. Lobo}
\email{fslobo@fc.ul.pt}
\affiliation{Instituto de Astrof\'{\i}sica e Ci\^{e}ncias do Espa\c{c}o, Faculdade de
Ci\^encias da Universidade de Lisboa, Edif\'{\i}cio C8, Campo Grande,
P-1749-016 Lisbon, Portugal}

\author{Giovanni Otalora}
\email{giovanni@ift.unesp.br}
\affiliation{Departamento de Matem\'atica, ICE, Universidade Federal de Juiz
de Fora, Minas Gerais, Brazil}\affiliation{Instituto de F\'isica Te\'orica, 
UNESP-Universidade 
Estadual Paulista
Caixa Postal 70532-2, 01156-970, S\~ao Paulo, Brazil}

\author{Emmanuel N. Saridakis}
\email{Emmanuel\_Saridakis@baylor.edu}
\affiliation{CASPER, Physics Department, Baylor University, Waco, TX 76798-7310, USA}
\affiliation{Instituto de F\'{\i}sica, Pontificia Universidad de Cat\'olica de 
Valpara\'{\i}so, 
Casilla 4950, Valpara\'{\i}so, Chile}

\date{\today}

\begin{abstract} 
In this work, we perform a detailed dynamical analysis for the cosmological applications of a nonminimal torsion-matter coupled gravity. Two alternative formalisms are proposed, which enables one to choose between the easier approach for a given problem, and furthermore, we analyze six specific models. In general, we extract fixed points corresponding either to dark-matter dominated, scaling decelerated solutions, or to dark-energy dominated accelerated solutions. Additionally, we find that there is a small parameter region in which the model can experience the transition from matter epoch to dark-energy era. These feature are in agreement with the observed universe evolution, and make the theory a successful candidate for the description of Nature.
\end{abstract}


\maketitle

\section{Introduction}\label{Introduction}

The late-time accelerated expansion of the universe is a major challenge to present-day cosmology. The converging observational evidence comes from a diverse set of cosmological data which includes observations of type Ia supernovae (SNeIa) \cite{PerlmutterandRiess}, cosmic microwave background radiation (CMB)  \cite{Spergel}, constraints from SDSS galaxy clustering \cite{Tegmark}, baryon acoustic oscillations (BAO) \cite{Eisenstein} and weak lensing  \cite{JainAndTaylor}. A plethora of theories have been proposed to explain this late-time cosmic acceleration. The simplest explanation is provided by a cosmological constant, however, this scenario is plagued by a severe fine-tuning problem associated with its energy scale \cite{Copeland:2006wr}. Hence one has two main directions in order to seek for an alternative explanation, namely, either to introduce the concept of an exotic repulsive cosmic fluid denoted by ``dark energy'' in the framework of general relativity \cite{Copeland:2006wr,Cai:2009zp}, or to modify the gravitational sector itself \cite{Nojiri:2006ri,Capozziello:2011et}.
All of these modifications to the Einstein-Hilbert action are based on the curvature description of gravity. However, an interesting and rich class of 
modified gravity can arise if one modifies the action of the equivalent torsional 
formulation of general relativity, known as ``Teleparallel Equivalent of General 
Relativity'' (TEGR) \cite{TEGR,JGPereira,Pereira:2012kd,Maluf:1994ji}. Thus, if one 
desires to modify gravity starting from this formulation, the simplest model is to 
extend the torsion scalar $T$ in the Lagrangian to an arbitrary function $f(T)$ 
\cite{Ferraro:2006jd,Linder:2010py,Chen:2010va,Tamanini:2012hg,Geng:2011aj,Wang:2011xf} 
(for a review see \cite{Cai:2015emx}).

Recently, an extension of $f(T)$ gravity was proposed in which the torsion scalar is coupled non-minimally to the matter sector \cite{NFT}. This theory is a novel gravitational modification, since it is different from both $f(T)$ gravity, as well as from the original nonminimal curvature-matter-coupled theory \cite{Bertolami:2007gv}. Additionally, the cosmological applications of this new theory proves to be very interesting \cite{NFT}. More specifically, a wide variety of cosmological scenarios were obtained, such as an effective dark energy sector whose equation-of-state parameter can be quintessence or phantom-like, or exhibit the phantom-divide crossing, while for a large range of the model parameters the Universe results in a de Sitter, dark-energy-dominated, accelerating phase. Furthermore, early-time inflationary solutions were also obtained, and thus these models provide a unified description of the cosmological history.
It is interesting to note, that along these lines of research, another extension of $f(T)$ gravity was also proposed \cite{Harko:2014aja} (motivated by Ref. \cite{Harko:2011kv}), that allowed for a general coupling of the torsion scalar $T$ with the trace of the matter energy-momentum tensor $\mathcal{T}$. The resulting $f(T,\mathcal{T})$ theory also possesses interesting cosmological phenomenology, leading to a unified description of the initial inflationary phase, the subsequent non-accelerating, matter-dominated expansion, and then the transition to a late-time accelerating phase. Additionally, a detailed study of the scalar perturbations at the linear level revealed that $f(T,\mathcal{T})$ cosmology may be free of ghosts and instabilities for a wide class of the model parameters.

In the present paper, we are interested in performing a detailed phase space analysis of the non-minimal torsion-matter coupling, and explore the phenomenology associated to some specific models. In particular, we apply the Dynamical System Approach, developed amongst others by Ellis and Wainwright \cite{Dynamical,ModDSA}. This approach is able to by-pass the non-linearities of the cosmological equations, that forbid an analytical treatment, and allow for the extraction of the general features and global behavior of the evolution, independently of the specific initial conditions. Using this technique we will be able to show that this class of models often present a matter-dominated epoch and that providing suitable forms for the functions of torsion, that appear in the theory, the universe evolves towards a late-time acceleration.

The manuscript is organised in the following manner: In Section, \ref{fTmodel} we briefly review the scenario of non-minimally torsion-matter coupled gravity and provide the corresponding cosmological equations. In Section, \ref{MSan1} we perform the dynamical analysis of this class of cosmologies, focusing on three different models. In Section \ref{MSan2}, for completeness, we present an alternative approach to the problem, which offers complementary information in relation to the first formulation. Finally, Section \ref{Conclusions} is devoted to the conclusions and discussion.

\section{The model}\label{fTmodel}

The action of non-minimally torsion-matter coupled gravity can be written in the following manner \cite{NFT}
\begin{equation}
S=\int{d^{4}x e \left[\kappa f_{1}(T)-f_{2}(T) \mathcal{L}_{m}\right]},
\label{Action}
\end{equation}
where $\kappa=1/16 \pi G$ and $f_{i}(T)$ (with $i=1,2$) are arbitrary functions of the torsion scalar $T$. The latter is defined as  
\begin{equation}
 T\equiv S_{\rho}^{~\mu\nu}\,T^{\rho}_{~\mu\nu}=\frac{1}{4}\,T_{\rho}^{~\mu 
\nu}\,T^{\rho}_{~\mu \nu}+\frac{1}{2}\,T^{\rho \mu}_{~~~\nu}\,T^{\nu}_{~\mu \rho}-T_{\rho 
\mu}^{~~\rho}T^{\nu\mu}_{~~\nu},
\end{equation} 
where
\begin{eqnarray}
 T^{\rho}_{~\mu\nu}\equiv 
e_{A}^{~\rho}\,\left(\partial_{\mu}e^{A}_{~\nu}-\partial_{\nu}e^{A}_{~\mu}
\right), \nonumber \\
  K^{\mu\nu}_{~~\rho}\equiv 
-\frac{1}{2}\left(T^{\mu\nu}_{~~\rho}-T^{\nu\mu}_{~~\rho}-T_{\rho}^{~\mu\nu}\right),
\nonumber\\
	S_{\rho}^{~\mu\nu}\equiv 
\frac{1}{2}\left(K^{\mu\nu}_{~~\rho}+\delta^{\mu}_{\rho}\,T^{\theta\nu}_{
~~\theta}-\delta^{\nu}_{\rho}\,T^{\theta\mu}_{~~\theta}\right),
 \label{torsion}
\end{eqnarray} 
are respectively  the torsion tensor, the contortion tensor and  the ``superpotential'' 
\cite{Cai:2015emx}.

Varying the action with respect to the tetrad $e^{A}_{\rho}$ yields the field
equations
\begin{eqnarray}
e^{-1} \partial_{\nu}{\left(e F e_{A}^{\rho} S_{\rho}^{~\mu\nu}\right)}+F e^{\tau}_{A} 
T^{\rho}_{
~\tau\nu} S_{\rho}^{~\nu\mu}+\frac{1}{4} \kappa f_{1} e^{\mu}_{A}\nonumber\\
+\kappa f_{2}' \partial_{\nu}{T} e^{\tau}_{A} \overset{\mathbf{em}}{S}_{\tau}{}^{\mu 
\nu}=\frac{1}
{4} f_{2} e^{\nu}_{A} \overset{\mathbf{em}}{T}_{\nu}{}^{\mu},
\label{geneoms}
\end{eqnarray}
where we have defined
\begin{equation}
\overset{\mathbf{em}}{S}_{A}{}^{\rho
\mu}=\frac{1}{4 \kappa} 
\frac{\partial{\mathcal{L}_{m}}}{\partial{\partial_{\mu}{e^{A}_{\rho}}}},
\label{Stilde}
\end{equation} 
and $F\equiv \kappa f_{1}'-f_{2}' \mathcal{L}_{m}$, with the prime denoting differentiation with respect to the torsion scalar. The matter energy-momentum 
tensor is defined as
\begin{equation}
\overset{\mathbf{em}}{T}_{A}{}^{\rho}=\frac{1}{e} 
\frac{\delta{S_{m}}}{\delta{e^{A}_{\rho}}}.
 \label{4}
\end{equation} In a purely space-time form, the Bianchi identities of Teleparallel Gravity 
imply 
the relationship
\begin{equation}
\bar{\nabla}_{\mu}\overset{\mathbf{em}}{T}_{\tau}{}^{\mu}=\frac{4}{f_{2}} K^{\rho}_{~\mu 
\tau} S_{\rho}^{~\mu\nu} 
\bar{\nabla}_{\nu}{F}-\frac{f_{2}'}{f_{2}}\left(\overset{\mathbf{em}}{T}_{\tau}{}^{\mu}
-\mathcal{L}_{m} \delta^{\mu}_{\tau}\right) \bar{\nabla}_{\mu}{T},
\label{Claw}
\end{equation} where $\bar{\nabla}_{\mu}$ is the covariant derivative in the Levi-Civita connection \cite{JGPereira}. Additionally, we have assumed that the matter Lagrangian $\mathcal{L}_{m}$,  depends only on the tetrad and not on its derivatives. Thus, the coupling between the matter and torsion describes an exchange of energy and momentum between both.

In order to obtain a cosmological model  governed by $f(T)$ gravity, we have
to impose the usual homogeneity and isotropy conditions. Therefore,
we consider the common choice for the tetrad field, given by
\begin{equation}
\label{weproudlyuse}
e_{\mu}^A={\rm
diag}(1,a(t),a(t),a(t)),
\end{equation}
which corresponds to a flat Friedmann-Robertson-Walker (FRW) background, with $a(t)$ the scale factor. Since the Lagrangian density of a perfect fluid is the energy scalar, representing the energy in a local rest frame for the fluid, a possible ``natural choice'' for the matter Lagrangian density is
$\mathcal{L}_{m}=\rho_{m}$ \cite{GroenHervik}. However, we mention that in the presence of non-minimal couplings in which the matter Lagrangian appears explicitly in the gravitational field equations, special care should be taken (we refer the reader to \cite{Bertolami:2008ab} for more details). Nevertheless, throughout this work we consider $\mathcal{L}_{m}=\rho_{m}$. In this case we have $\overset{\mathbf{em}}{S}_{A}{}^{\rho \mu}=0$, and also the usual 
form of the energy-momentum tensor for the perfect fluid is given by
\begin{equation}
\overset{\mathbf{em}}{T}_{\mu\nu}=(\rho_{m}+p_{m}) u_{\mu} u_{\nu}-p_{m} g_{\mu
\nu}. 
\end{equation}
One can see that the energy-momentum tensor is again
conserved, just like in teleparallel gravity or $f(T)$ theories, since Eq. \eqref{Claw}
yields the continuity equation 
\begin{equation}
\dot{\rho}_{m}+3 H \left(1+\omega_{m}\right)\rho_{m}=0,
\label{Ceq}
\end{equation} 
where $H=\dot{a}/a$ is the Hubble parameter and $\omega_{m}\equiv 
p_{m}/\rho_{m}$ is the equation-of-state (EOS) parameter of matter. In the above expressions the dot represents a derivatives with respect to cosmic time.

Inserting the flat FRW vierbein choice (\ref{weproudlyuse}) into the
field equations \eqref{geneoms} we obtain the modified Friedmann equations \cite{NFT}
\begin{eqnarray}
\label{H0}
 12 F H^2&=&f_{2}\rho_{m}-\kappa f_{1},\\
\label{H00}
\dot{F} H+F \dot{H}&=&-\frac{1}{4}\gamma f_{2} \rho_{m},
\end{eqnarray}
where we have defined $\gamma\equiv 1+\omega_{m}$. Note that a useful 
relation arising from the choice \rf{weproudlyuse} of the tetrad is that the 
torsion scalar is proportional to the square of the Hubble parameter, namely $T=-6H^2$.

The generalized Friedmann equations can be rewritten in the standard form 
\begin{eqnarray}
  3H^2&=& \frac{1}{2\kappa}\left(\rho_{DE}+\rho_m  \right), \label{Fr1} \\
  2\dot{H}& =&-\frac{1}{2\kappa}\left(\rho_{m}+p_{m}+\rho_{DE}+p_{DE}\right), 
\label{Fr2}
\end{eqnarray}
where the effective energy density and effective pressure of the dark energy sector are 
defined as
\begin{equation}
\label{rhode}
 \rho_{DE}=\left(\frac{\kappa f_{2}}{2 F}-1\right)\rho_{m}-\frac{\kappa^2 f_{1}}{2 F},
\end{equation}
\begin{equation}\label{pde}
p_{DE}=\left(\frac{\kappa f_{2}}{F}-1\right)\gamma\rho_{m}+\frac{4\kappa H 
\dot{F}}{F}-\rho_{DE},
\end{equation}
respectively. One can easily verify that the dark energy density and pressure satisfy the continuity equation
\begin{eqnarray}\label{cons}
 \dot{\rho}_{DE}+3H\left(1+\omega_{DE}\right)\rho_{DE}=0,
\end{eqnarray} in accordance with Eqs. \eqref{Claw} and \eqref{Ceq}. We have defined the dark-energy equation-of-state parameter as usual, namely,
$\omega_{DE}\equiv p_{DE}/\rho_{DE}$.

\section{Dynamical system analysis}\label{MSan1}

In order to perform the phase-space analysis of the cosmological scenario at hand, we have to introduce suitable dimensionless auxiliary variables that will bring the system of cosmological equations into its autonomous form \cite{Dynamical,ModDSA}. We choose
\begin{equation}
\label{VarS}
 X=\frac{f_1}{12 H^2 f_1'},\quad Y=\frac{f_2}{12 H^2 f_2'},\quad\Omega =\frac{\rho _m}{6 
H^2 
\kappa },
\end{equation}
and the logarithmic time $N=\log a$. With this choice, and taking in account   that 
$T=-6 H^2$, the case of a standard $f(T)$ theory is recovered when 
$|Y|\rightarrow\infty$. 
 
Using the variables \rf{VarS} the cosmological equations  are equivalent to the autonomous 
system
\begin{equation}
\begin{split}\label{dynsysS}
&\DerN X=-\frac{3 \gamma  {\bf Q} (X+1) (Y+1) (2 ({\bf W}+1) X+1)}{{\bf P} (X+1)-{\bf Q} 
(2 {\bf W} 
(Y+1)-X+Y)},\\
&\DerN Y=-\frac{3 \gamma  (X+1) (Y+1) ({\bf P} Y+2 {\bf Q} Y+{\bf Q})}{{\bf P} (X+1)-{\bf 
Q} (2 {\bf W} (Y+1)-X+Y)},\\
&2 {\bf Q} (Y+1) \Omega +X+1=0,
\end{split}
\end{equation}
with $\gamma\equiv 1+\omega_{m}$, as defined above, and
 where 
 \begin{equation}
\begin{split}
{\bf Q} =\frac{T f_2'}{2 f_1'},\qquad {\bf P}=\frac{T^2 f_2''}{f_1'}, \qquad {\bf W}=\frac{T f_1''}{f_1'}\,,
\end{split}
\label{VarSS}
\end{equation}
are functions only of $T$. Their expression in terms of the variables \rf{VarS} can be obtained noting that 
\begin{equation}
\frac{X}{Y}=\frac{f_1 f_2'}{f_2 f_1'}.
\end{equation}
Inverting the above equation for a given form of $f_1$ and $f_2$, one obtains $T=T(X/Y)$ and the system \rf{dynsysS} can be closed. It is important to stress at this point that since $T=-6 H^2$ must be non positive, the requirement that $T(X/Y)\leq0$ implies that only some parts of the phase space defined by those variables will have actual physical meaning. In fact, this restriction implies that even if some physically interesting fixed points  are present  there might not be physical orbits that connect them. Additionally, note that since we have used the third equation of \rf{dynsysS} to eliminate $\Omega$, we might need to discard the $Y=-1$ part of the phase space. 

Using the dimensionless variables from Eqs. \eqref{VarS} and \eqref{VarSS}, the  
dark-energy density parameter is written as 
\begin{equation}
\Omega_{DE}\equiv \frac{ \rho_{DE}}{6\kappa H^2}=1-\Omega,
\end{equation}
Due to the definition \rf{VarS} the physically relevant part of the phase space 
will necessarily have $\Omega\geq 0$, however, since $\Omega_{DE}$ does not need to be defined positive, $\Omega$ is not necessarily less than one. In the following however, we will require, at least for the matter-dominated fixed points, that $0<\Omega< 1$ and therefore that  $0<\Omega_{DE}<1$. 
 
The solutions associated to the fixed points can be obtained using the equation
\begin{equation}\label{Ray1S}
\begin{split}
&\frac{\dot{H}}{H^2}= -\frac{1}{p},\\
& \frac{1}{p}=-\frac{3 \gamma  {\bf Q}_* (X_*+1) (Y_*+1)}{{\bf P}_* (X_*+1)-{\bf Q}_* (2 
{\bf W}_* (
Y_*+1)-X_*+Y_*)},
\end{split}
\end{equation} 
where $p$ is a constant and an asterisk represents the value of a variable at the 
fixed point $\{X_* $,$Y_*$\}. Hence, the solutions obtained in this way are all scaling solutions which can alleviate the coincidence problem \cite{Copeland:2006wr}.

The parameter of deceleration is defined as 
\begin{equation}
q\equiv -\frac{\dot{H}}{H^{2}}-1=\frac{1}{p}-1.
\end{equation}
Thus, when $H=H_0={\rm constant}$ (and therefore $q=-1$) we have the de Sitter solution $a(t)\propto e^{H_0 t}$ and accelerated expansion for all values. For finite $p>0$ we acquire $a(t)\propto t^{p}$, and we have an accelerated solution if $p>1$ ($q<0$) or a decelerated solution if $0<p<1$ ($q>0$).  

In the following subsections we apply the above general technique to three 
specific cosmological models, obtained by specific choices of the functions $f_1$ and $f_2$.

\subsection{Model I: $f_{1}(T)=T+\alpha T^2-\Lambda$ and $f_{2}(T)=1+\beta T$}

As a first example we consider the specific choices, given by 
\begin{equation}\label{Model1}
f_{1}(T)=T+\alpha T^2-\Lambda,\:\:\:\text{and}\:\:\: f_{2}(T)=1+\beta T,
\end{equation}with $\alpha$, $\beta$ and $\Lambda$ constants. In this case, the 
auxiliary variable choice \rf{VarS} implies
\begin{equation}
\frac{X}{Y}=\frac{\beta  \left(-\Lambda +\alpha  T^2+T\right)}{(2 \alpha  T+1) (\beta  
T+1)},
\label{TModelI}
\end{equation}
and
\begin{align}\label{Tcase1}
&T(X,Y)=\frac{\beta  Y-(2 \alpha  +\beta)  X + K}{2 \alpha  \beta (2  X-Y)},\\
&K=\pm\sqrt{[(2 \alpha+\beta)  X-\beta  Y]^2-4 \alpha \beta (2 X-Y) (X+\beta  \Lambda  
Y)}\,.
\end{align}
Therefore, for these choices of the functions $f_1$ and $f_2$, we have two possible 
solutions for $T(X/Y)$. In the following, we will choose the solution with positive $K$. Inserting the functions in Eq. \rf{VarSS}, one has 
\begin{equation}
{\bf Q} =\frac{\beta  T}{2\left(1+2 \alpha  T\right)}, \quad {\bf P}=0, \quad {\bf 
W}=\frac{2 \alpha  T}{2 \alpha  T+1}.
\end{equation}

We will analyse the autonomous system and the fixed points, and their respective stability below.

\def\tablename{Table}%
\begin{table*}[ht!]
\centering
\begin{center}
\begin{tabular}{c c c c c c c}\hline\hline 
Name         &  $\left\{X_{*},Y_{*}\right\}$   & \ \ $a(t)$ & \ \ $\rho(t)$ 
&$\Omega_{DE}=1-\Omega$ & 
Deceleration parameter $(q)$ & \ Stability \\ \hline \\
$\mathcal{A}$ & $\left\{-1, Y_*\right\}$ & $e^{H_0 t}$ & 0  &$1$ & $-1$ & Attractors\\ \\
$\mathcal{B}$ & $\left\{X_0, -\frac{1}{2}\right\}$  &  $t^{p}$ & $t^{-3\gamma p}$ & 
$1-\frac{16 \alpha  X_0 \left(X_0+1\right)}{\beta +2 X_0 (2 \alpha +\beta )-4 \alpha  
X_0}$  & $\frac{1}{p}-1$& 
Saddle\\ \\ \hline\\
\multicolumn{7}{c}{$H_0=\frac{1}{6} \sqrt{\frac{1+\sqrt{1-12 \alpha  \Lambda }}{\alpha }}, 
\quad X_
0= \frac{4 \alpha  \beta  \Lambda +\beta -2 \sqrt{4 \alpha ^2 \beta ^2 \Lambda ^2+4 \alpha 
^2 \beta 
 \Lambda +\alpha \beta ^2 \Lambda +\alpha  \beta }}{2 (4 \alpha -\beta )}, \quad 
p=\pm\frac{4 
\alpha}{
3 \gamma}\sqrt{  \frac{\beta  \Lambda +1}{\alpha  \beta  (4 \alpha  \Lambda +1)}}$}\\
 \\\hline\hline
\end{tabular}
\end{center}
\caption{The critical points for Model I. Here $Y_*$ are the values of Y for which 
$T=-6H^2<0$.}
\label{table1}
\end{table*}

\subsubsection{Autonomous system and fixed points}

Substituting the expressions for ${\bf Q}$ , ${\bf P}$ and  ${\bf W} $ into the dynamical 
system \eqref{dynsysS} we find 
\begin{equation}
\begin{split}\label{example 2a}
&\DerN X=-\frac{3 \gamma  (X+1) (Y+1) \left[2 \alpha  T+ 2 X (4 \alpha  T+1)+1\right]}{2 
\alpha  T (
X-3 Y-2)+X-Y},\\
&\DerN Y=-\frac{3 \gamma  (X+1) (Y+1) (2 Y+1) (2 \alpha  T+1)}{2 \alpha  T (X-3 Y-2)+X-Y},
\end{split}
\end{equation}
where $T$ is given by \rf{Tcase1}. The dynamical system presents the invariant 
submanidfolds $X=-1$, $Y=-1$, $Y=-1/2$, $T=-1/2\alpha$. Some of these submanifolds, such as $X=-1$, represent fixed points, whereas others can be singular, as in the case of $Y=-1$. The presence of invariant submanifolds additionally implies that the phase space does not admit a global attractor.

The critical points are obtained  setting $dX/dN = dY/dN = 0$ and are shown in Table \ref{table1}. They consist of a line of fixed points $\mathcal{A}$ characterized by $X=-1$, associated via \rf{Ray1S} to the solution $a(t)\propto e^{H_0 t}$ with
\begin{equation}
H_0=\frac{1}{6} \sqrt{\frac{1+\sqrt{1-12 \alpha  \Lambda }}{\alpha }},
\end{equation} 
and $\Omega\propto \rho_m=0$, i.e., they correspond to a dark-energy dominated, de Sitter universe. 

In addition to the line $\mathcal{A}$, the phase space admits the couple of fixed 
points  $\mathcal{B}$ for $Y=-1/2$. They are associated to the solution  $a(t)\propto t^{p}$ with $H=p/t$ and 
\begin{equation}
\label{P_B}
p=\pm\frac{4 \alpha}{3 \gamma}\sqrt{  \frac{\beta  \Lambda +1}{\alpha  \beta  (4 \alpha  
\Lambda +1)
}},
\end{equation} 
where the $\pm$ signs guarantee that $p>0$. 

Points $\mathcal{B}$ exist only if the inequality $\alpha X_\mathcal{B}<0$ is satisfied, and they correspond to physical states only if the torsion scalar evaluated in their coordinates is negative. The same happens with the line $\mathcal{A}$: in general, only segments of it will be in the physical $T<0$ part of the phase space. These complex constraints make a complete analysis of this example, which are too complicated to be fully included here. Hence, in order to highlight physically interesting scenarios we will choose parameter values such that at least one of the points $\mathcal{B}$ is in the $T<0$, $1<\Omega<0$ region of the phase space. One set of parameter values for which this is possible is
\begin{align}
&\alpha >0, \qquad \beta>2\alpha, \qquad \frac{\beta -16 \alpha }{12 \alpha  \beta }<\Lambda <\Lambda_* ,
\\
\nonumber &\Lambda_*=-\frac{64 \alpha ^3+24 \alpha ^2 \beta +\beta ^3}{32 \alpha ^2 \beta  
(3 \alpha +\beta
   )}
-\frac{\sqrt{16 \alpha ^2+\beta ^2} }{32 \alpha ^2}\left| \frac{\beta ^2-16 \alpha 
^2}{\beta(3 \alpha +\beta )}\right| .
\end{align}
With these choices, $\mathcal{A}$ satisfies the constraint on $T<0$ for
\begin{equation}
\frac{1}{\beta
    \Lambda }<Y\leq 2 \sqrt{\frac{-\alpha ^2+\alpha  \beta ^2 \Lambda +\alpha  \beta
   }{\beta ^2 (4 \alpha  \Lambda +1)}}-1,
\end{equation}
and moreover only one point $\mathcal{B}$ exists  for which $0<\Omega<1$  and  $0<p<1$.

\subsubsection{Stability of fixed points}

The stability of the fixed points is deduced using the Hartmann-Grobmann theorem 
\cite{MartinGuterman}, and the stability properties for each critical point are shown in 
Table \ref{table1}. The de Sitter line $\mathcal{A}$ is always attractive with eigenvalues
\begin{equation}
\mu_{1}=0,\:\:\:\: \mu_{2}=-3\gamma.
\end{equation} 
Here the zero eigenvalue is related to the fact that we are dealing with a line of fixed 
points rather than with a non-hyperbolic character of the point. The character of the 
line is determined by the sign on the non-zero eigenvalue and is therefore attractive. 

\begin{figure}[ht!]
\includegraphics[width=0.45\textwidth]{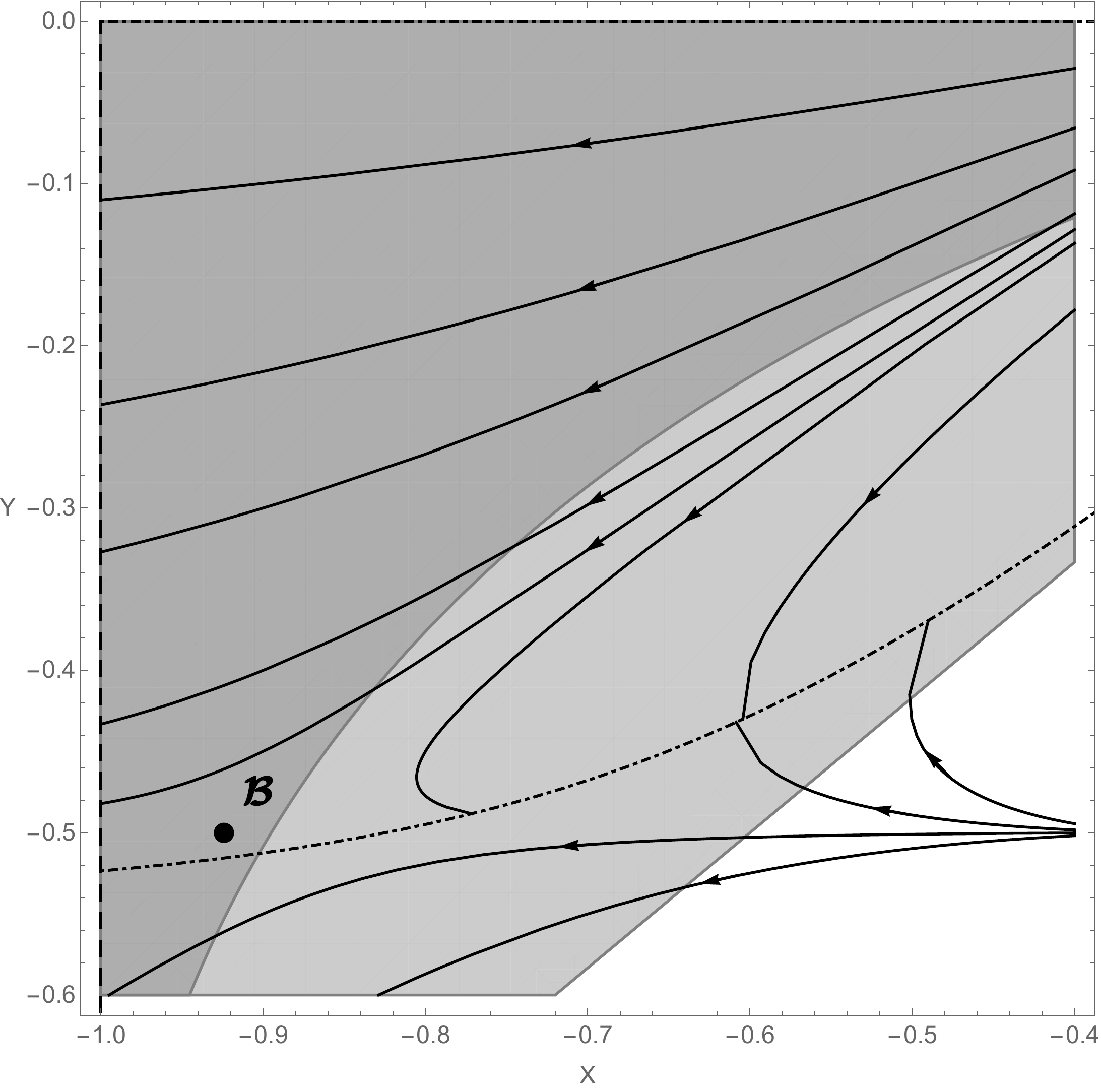}
\caption{{\it {An example of the phase space for Model I: $f_{1}(T)=T+\alpha 
T^2-\Lambda$ and $f_{2}(T)=1+\beta T$, with $\alpha=1, 
\beta=1, \Lambda=-6/5$. The  dark grey area is the part of the phase space for which $T<0$ and 
the light gray one is the one in which $0<\Omega<1$.  The dashed line represents the line of fixed points 
$\mathcal{A}$. The 
dot-dashed lines represent the value of the phase space for which the dynamical equations are 
singular.}}}
\label{FigureModel1}
\end{figure}

On the other hand, the eigenvalues of the critical points $\mathcal{B}$ are given by
\begin{equation}
\mu_{1}=\frac{3}{2} \gamma  \sqrt{ \frac{\beta (4 \alpha  \Lambda +1)}{ \alpha (\beta  
\Lambda +1)}}, \qquad
 \mu_{2}=3\gamma  \left(2-\frac{ \beta }{2 \alpha }\right).
\end{equation} 
It follows that point $\mathcal{B}$ can only be a saddle in the parameter range we consider. 
Choosing suitable initial conditions, it is therefore possible that the model will present 
an unstable phase of decelerated expansion followed by accelerated expansion. An example 
of this case is shown in the phase-space portrait of Fig. \ref{FigureModel1}, 
arising through a numerical elaboration in the case of $\alpha=1, 
\beta=1, \Lambda=-6/5$. The phase space presented also reveals 
that there are orbits which evolve towards the singularity of the system. This could 
deduce an instability for this class of models, and therefore it restricts the range of 
viable initial conditions.

\subsection{Model II: $f_{1}(T)=T-\Lambda$ and $f_{2}(T)=1-\alpha 
T+\beta T^2$}

We consider now the case 
\begin{equation}\label{Model2}
f_{1}(T)=T-\Lambda,\:\:\:\text{and}\:\:\: f_{2}(T)=1-\alpha T+\beta T^2,
\end{equation}
where $\alpha$, $\beta$ and $\Lambda$ are constants.
For this model we have that
\begin{equation}
\frac{X}{Y}=\frac{(T-\Lambda ) (2 \beta  T-\alpha)}{T (\beta  T-\alpha )+1},
\label{TModelII}
\end{equation}
and therefore 
\begin{align}\label{Tcase2}
&T(X,Y)= \frac{\alpha  X-Y (\alpha +2 \beta  \Lambda )}{2 \beta  (X-2 Y)},\\
&K=\pm\sqrt{[\alpha  X-Y (\alpha +2 \beta  \Lambda )]^2-4 \beta  (X-2 Y) (X-\alpha  
\Lambda Y)} .
\end{align}
Thus, similarly to the Model I above, in this case there are multiple solutions for 
the expression of the torsion in terms of the variables $X, Y$. In the following we will 
choose the expression where $K$ is positive.

The parameters ${\bf Q} , {\bf P}, {\bf W}$ become
\begin{equation}
\begin{split}
{\bf Q} =-\frac{1}{2} T (\alpha -2 \beta  T),\qquad 
 {\bf P}=2 \beta  T^2, \qquad
 {\bf W}=0.
\end{split}
\end{equation}
The conditions $T\leq0$  and $0\leq\Omega\leq 1$ lead to complex constraints of the 
allowed intervals of the variables, which depend on the values of the constants $\alpha, 
\beta, \Lambda$. In order to select a class of physically interesting cases, we choose 
\begin{equation}
\alpha>0, \qquad \beta<0,\qquad \frac{\alpha }{4 \beta }<\Lambda<\Lambda_*,
\end{equation}
where $\Lambda_*$ is a solution of the algebraic equation 
\begin{align}
\nonumber &\alpha ^5-3 \alpha ^3 \beta -\Lambda ^2 \left(-16 \alpha ^5 \beta +112 \alpha 
^3 \beta
   ^2-240 \alpha  \beta ^3\right)\\
&-\Lambda  \left(\alpha ^6+12 \alpha ^4 \beta -100 \alpha
   ^2 \beta ^2+192 \beta ^3\right)-64 \beta ^4 \Lambda ^3 =0.
\end{align}

\def\tablename{Table}%
\begin{table*}[ht!]
\centering
\begin{center}
\begin{tabular}{c c c c c c c}\hline\hline
Name &  $\left\{X_{*},Y_{*}\right\}$ &\ \  $a(t)$ & \ \ $\rho(t)$  & 
$\Omega_{DE}=1-\Omega$  & 
Deceleration 
factor $(q)$  & \ Stability \\\hline \\ 
$\mathcal{A}$ & $\left\{-1,Y_*\right\}$& $e^{H_0 t}$ & 0 & $1$ & $-1$ & Attractor \\ \\ 
$\mathcal{B}$ & $\left\{-\frac{1}{2},Y_0\right\}$  &$t^{p}$ & $t^{-3\gamma p}$ & 
$1+\frac{\beta  \left(4 Y_0+1\right){}^2}{2 \alpha  Y_0 \left(Y_0+1\right) \left(\alpha +2
   \alpha  Y_0\right)}$
   & $\frac{1}{p}-1$ & Repeller\\ \\  \hline \\
\multicolumn{7}{c}{$H_0=\sqrt{\frac{\Lambda}{6}}, \quad Y_0=\frac{4 \beta -\alpha ^2- 
2\sqrt{\beta \left( \alpha ^2-4 \beta\right) (\alpha  \Lambda -1)}}{2 \alpha  (\alpha +4 
\beta  \Lambda )}, \quad 
 p=\frac{4  \sqrt{\beta(\alpha  \Lambda -1)}}{3 \gamma \sqrt{\alpha ^2-4 \beta }} $}\\ \\
\hline\hline
\end{tabular}
\end{center}
\caption{The critical points for Model II in the case $\beta>\alpha^2/4$. Here $Y_*$ are 
the values of Y for which $T=-6H^2<0$.} 
\label{Table3}
\end{table*}

\subsubsection{Autonomous system and fixed points}

Substituting the expressions for ${\bf Q} ,{\bf P} ,{\bf W} $, the dynamical equations  
become
\begin{equation}
\begin{split}\label{example 2}
&\DerN X=-\frac{3 \gamma  (X+1) (2 X+1) (Y+1) (\alpha -2 \beta  T)}{\alpha  (X-Y)-2 \beta  
T (3 X-
Y+2)},\\
&\DerN Y=-\frac{3 \gamma  (X+1) (Y+1) \left[(1+ 2 Y)\alpha-2 \beta  T (4 Y+1)  
\right]}{\alpha  (X-
Y)-2 \beta  T (3 X-Y+2)},
\end{split}
\end{equation} 
where $T$ is given by \rf{Tcase2}. The system is similar to the one of 
Mode I above, and presents the same invariant submanifolds. Setting to zero 
the L.H.S. of the equations we find a critical line $\mathcal{A}$ for $X=-1$. In the 
chosen interval for the parameters this line is physical ($T<0$) for $\alpha<0$ and 
\begin{align}
 &Y\leq Y_1 \quad \land \quad Y>-1/2,\\
 & \nonumber Y_{1}= \frac{4 \beta -\alpha ^2}{(\alpha +2 \beta  \Lambda )^2}
 -2 \sqrt{-\frac{\beta  \left(\alpha ^2-4 \beta \right) (\Lambda  (\alpha 
+\beta  \Lambda
   )+1)}{(\alpha +2 \beta  \Lambda )^4}}.
\end{align}
Furthermore, the system admits two additional fixed points. However, in the parameter 
region we have considered, only one of them, named $\mathcal{B}$, lies in the $T<0$ part 
of the phase space.

The solutions associated to these fixed points are obtained from Eq. \rf{Ray1S}. The 
points on the critical line $\mathcal{A}$ satisfiy $a\sim e^{H_0 t}$ with 
\begin{equation}
H_0=\sqrt{\frac{\Lambda}{6}}={\rm const},
\end{equation} 
and thus they represent an accelerated expansion for all parameter values, with 
$q=-1$ and $\Omega=0$,  i.e., they correspond to a dark-energy dominated, de-Sitter 
universe. On the other hand, point $\mathcal{B}$ represents the power-law 
solutions  $a\sim t^{p}$ with $H=p/t$ and
\begin{equation}\label{P_B2}
p=\frac{4 \sqrt{\beta(\alpha  \Lambda -1)}}{3 \gamma \sqrt{\alpha ^2-4 \beta }}.
\end{equation} 
Note that with our choice of parameters, $\alpha  \Lambda -1$ is always 
positive, and therefore this solution always represents a decelerated solution (with 
$0\leq\Omega\leq 1$). 

\subsubsection{Stability of fixed points}
 
\begin{figure}[ht!]
\includegraphics[width=0.45\textwidth]{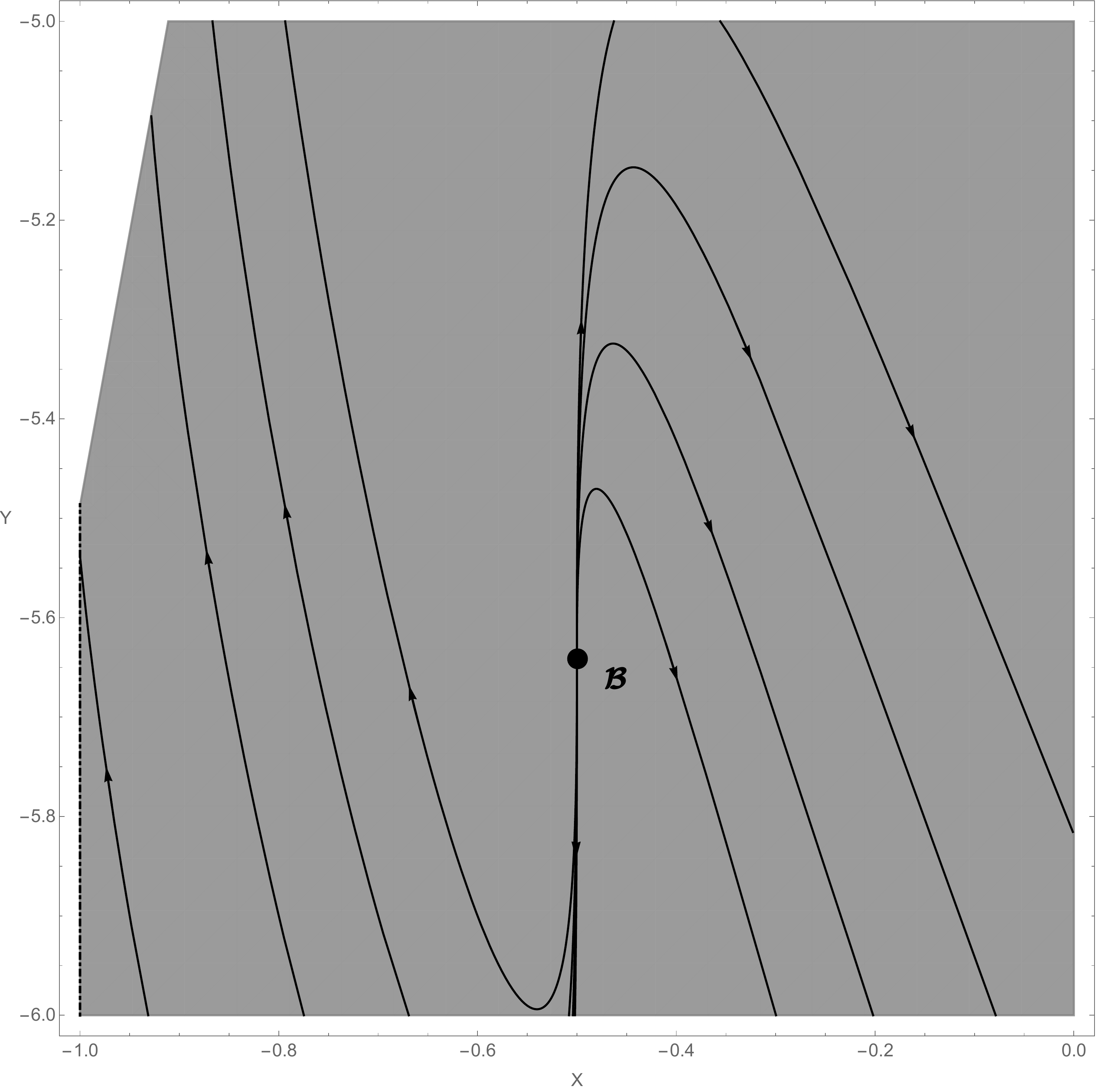}
\caption{{\it {An example of the phase space for Model II: $f_{1}(T)=T-\Lambda$ and 
$f_{2}(T)=1-\alpha 
T+\beta T^2$, with $\alpha=5, 
\beta=-18/5, \Lambda=-2/7$. The grey area is the part of the phase space for which $T<0$ 
and $0<\Omega<1$.  The 
dot-dashed line represents the line of fixed points $\mathcal{A}$ that have $T<0$.}}}
\label{FigureModel2}
\end{figure}

The de Sitter line $\mathcal{A}$ is always attractive with eigenvalues $\mu_{1}=0$ and 
$\mu_{2}=-3\gamma$. On the other hand, for the critical point $\mathcal{B}$ the 
eigenvalues are given by
\begin{equation}
\mu_{1}=\frac{3 \gamma  \sqrt{\alpha ^2-4 \beta }}{2  \sqrt{\beta(\alpha  \Lambda 
-1)}},   \qquad 
\mu_2=-\frac{3 \gamma  (\alpha -4 \beta  \Lambda )}{2 \beta  \Lambda }.
\end{equation}
Hence, for the physically meaningful region $0\leq\Omega\leq1$ and $T\leq0$, the critical 
point $\mathcal{B}$ is always a repeller and only for very specific initial conditions there are orbit that can connect $\mathcal{B}$ to the line of fixed points $\mathcal{A}$ within the physical part of the phase space. 
The critical points of Model II and their features, are summarized in Table \ref{Table3}. Finally, for completeness, in 
Fig. \ref{FigureModel2} we provide a specific phase-space portrait, arising from a 
numerical analysis for the case of $\alpha=5, 
\beta=-18/5, \Lambda=-2/7$.  

\subsection{Model III: $f_{1}(T)=G e^{A T^{2}}$ and $f_{2}(T)=\Lambda e^{- 
T}$}

We consider now the case with exponential couplings
\begin{equation}\label{Model3}
f_{1}(T)=G e^{A T^{2}},\:\:\:\text{and}\:\:\: f_{2}(T)=\Lambda e^{-T},
\end{equation} where  $A$, $G$ and $\Lambda$ are constants.

For the form of coupling given above  we have that
\begin{equation}
T=-\frac{Y}{2A X},
\label{Tcase3}
\end{equation}
and
\begin{eqnarray}
&&{\bf Q} =-\frac{\Lambda}{4 A G} e^{\frac{Y (2 X-Y)}{4 A X^2}},\\ \nonumber
&&{\bf P}=\-\frac{\Lambda  Y }{4 A^2 G X}e^{\frac{Y
   (2 X-Y)}{4 A X^2}}, \\ \nonumber
&&{\bf W}=\frac{Y^2}{2 A X^2}+1.
\end{eqnarray}

Hence, in this case only $Y/AX>0$ represents the physical part of the phase space. In the 
following, for simplicity, we will only consider $\Lambda>0$, 
$0<G<3/7$ and $A>0$.   
\def\tablename{Table}%

\subsubsection{Autonomous system and fixed points}

For this model the dynamical equations are written as 
\begin{eqnarray} \label{ModelII}
&& \DerN X=\-\frac{3 \gamma X (X+1) (Y+1) \left[A X (4 X+1)+Y^2\right]}{A X^2 (X-3 Y-2)+Y
   (X-Y) (X+Y+1)},\\
&& \DerN Y=\frac{3\gamma X (X+1) (Y+1) \left[A (2 X Y+X)+Y^2\right]}{A X^2
   (2+3 Y-X)+Y (Y-X) (X+Y+1)},
\end{eqnarray}
As in the previous models, the above system presents the invariant submanifolds $X=0$, $X=-1$ and $Y=-1$ and their existence implies that no global attractor exists in the phase space. The submanifold $X=0$ has to be excluded as it represents a divergent torsion and in order to use our  general approach one has to exclude  $Y=-1$.

The system admits, therefore, a line of fixed points and an isolated fixed point.  The critical line $\mathcal{A}$ is a de-Sitter 
fixed point such that $a\sim e^{H_0 t}$ with 
\begin{equation}
H_0=\frac{1}{2 \sqrt{3} \sqrt[4]{A}},
\end{equation}
and corresponds to $T<0$ only if $Y_{*}<0$. The critical point $\mathcal{B}$ is a scaling 
solution with $X_{*}/AY_{*}=1$. The associated solution is obtained from 
\rf{Ray1S}, and it represents a power-law solution $a\sim t^p$ with $H=p/t$ and exponent $p$ given by
\begin{equation}
p=\frac{2 (A+1)}{3 \gamma}.
\end{equation}  
Thus, it can be either an accelerated or decelerated solution.

\subsubsection{Stability of fixed points}

As in the previous examples, using the Hartman-Grobman theorem we can analyze the 
stability of the 
fixed subspaces.  For the line $\mathcal{A}$ the eigenvalues are 
\begin{equation}
\mu_{1}=0, \qquad  \mu_{2}=-3 \gamma.
\end{equation} 
which is independent for $Y_*$. For point $\mathcal{B}$ they read 
\begin{eqnarray}
\mu_{1,2}=\frac{3 \gamma  \left(3 A\pm \sqrt{(A-8) A}\right)}{2 (A+1)}
\end{eqnarray} 
so that for $A>0$ this point can be a repeller or a saddle. 
The critical points of Model III and their features, are summarized in Table \ref{Table5}.

\begin{figure}
\centering
\includegraphics[width=.95\linewidth]{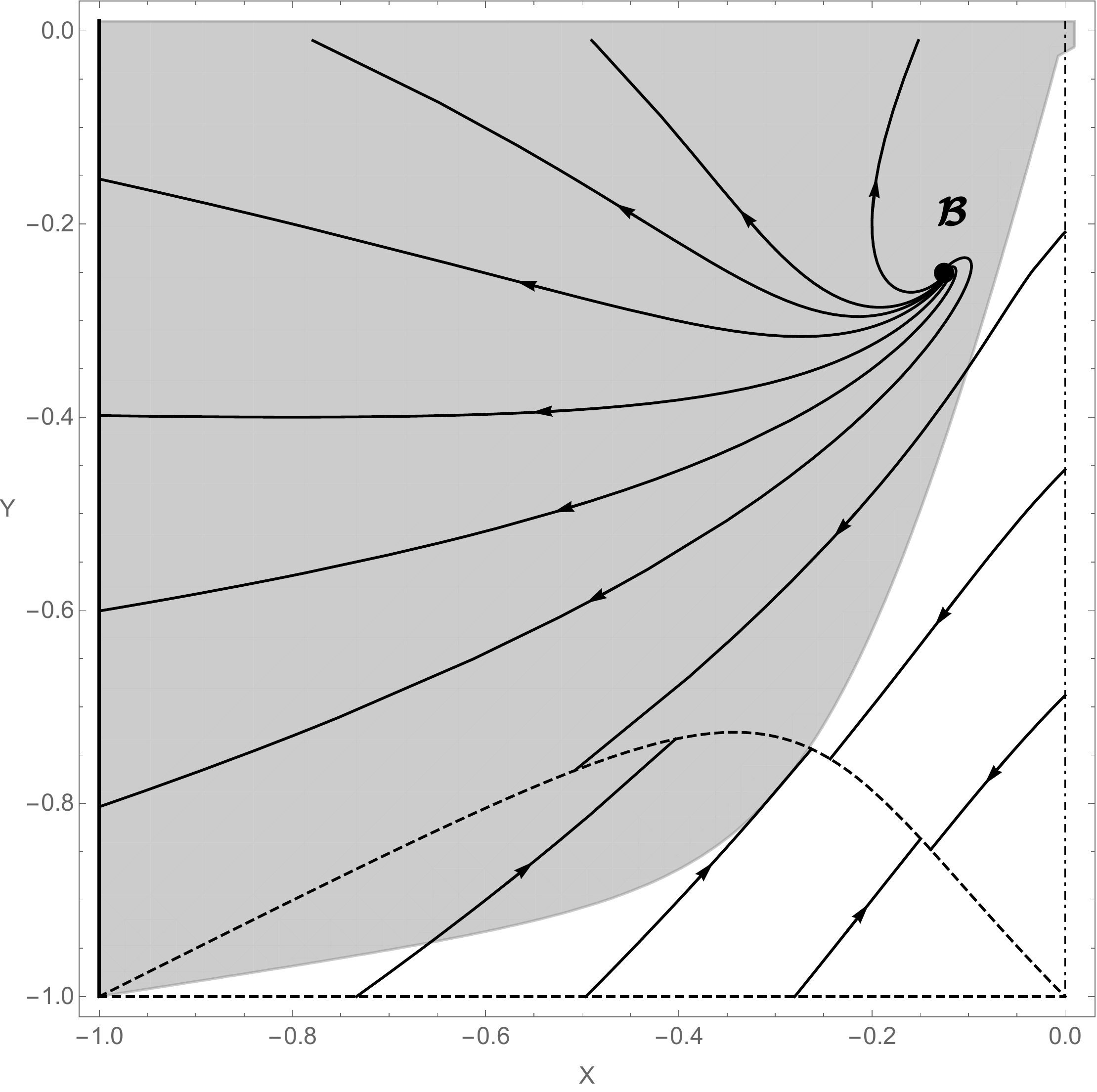}
\caption{{\it {An example of the phase space of Model III: $f_{1}(T)=\Lambda e^{A T}$ 
and $f_{2}(T)=e^{\frac{A}{2 \alpha} T^{-2 \alpha }}$, with $ A= 1, \gamma=1, \Lambda=1, G= 1/10$. 
Here the grey area represents the region in which $0<\Omega<1$, the balck line is the fixed line 
$\mathcal A$, the dot -dashed line in the invariant submanifold $X=0$, the dashed lines 
represent singularities for the dynamical system.}}}
\label{FigModel3}
\end{figure}

It turns out that this model is able, for the interval of parameters we have considered, to describe 
a transition between decelerated and accelerated expansion. In Fig. \ref{FigModel3}, we present a phase-space portrait for $ A= 1, \gamma=1, \Lambda=1, G= 1/10$. Similarly to Model I above, in this case there also appear 
orbits which evolve towards singular states for the model.
%


\begin{table*}[ht!]
\centering
\begin{center}
\begin{tabular}{c c c c c c c}\hline\hline
Name &  $\left\{X_{*},Y_{*}\right\}$ & $a(t)$ & $\rho(t)$ & $\Omega_{DE}=1-\Omega$  &  
Deceleration 
factor $(q)$  & Stability\\ \\\hline \\
$\mathcal{A}$ & $\left\{-1, Y_{*}<0\right\}$&$e^{H_0 t}$ & 0 & $1$ &$-1$ & Attractor \\ \\
$\mathcal{B}$ & $\left\{-\frac{A}{4 (A+1)},-\frac{A}{2 (A+1)}\right\} $  &$t^{\frac{2 (A+1)}{3 \gamma}}$ & 
$t^{-2 (A+1)}$ & $1-\frac{A (3 A+4) G}{(A+2) \Lambda }$ 
&$\frac{3 A\gamma}{2(1+A)}-1$ & Saddle for $A>8$\\
&&&&&& Repeller (focus) for $0<A<8$\\
\hline\hline
\end{tabular}
\end{center}
\caption{The critical points for Model III.} 
\label{Table5}
\end{table*}

\section{An alternative dynamical-systems approach}\label{MSan2}

The analysis performed in the previous section does not make full use of the key 
relationship $T=-6H^2$. In order to formulate an approach which will be optimal in this 
respect, let us consider the form of Eq. \rf{H0}. Since $\mathcal{L}_{m}=\rho_{m}$  one has
\begin{equation}\label{H0_1}
\kappa  \left(12 H^2 f_1'+f_1\right)-\rho _m \left(12 H^2 f_2'+f_2\right)=0.
\end{equation}
Using the relation $T=-6 H^2$, this equation can be rewritten as 
\begin{equation}\label{H0_21}
 \frac{f_1(H^2)+12 H^2  f_1'(H^2)}{f_2(H^2)+12 H^2  f_2'(H^2)}- \frac{\rho _m}{\kappa }=0.
\end{equation}
In the above equation the first term is a function of $H^2$ only, and hence it could 
be formally solved for $H^2$ giving
\begin{equation}
\label{H0_2}
 H^2 - g\left(\frac{\rho _m}{\kappa }\right)=0,
\end{equation}
where $g\left(\rho _m/\kappa \right)$ a function of its argument. Relation 
(\ref{H0_2}) implies that this scenario is equivalent to general relativity plus a fluid 
with a non-trivial equation of state. Additionally, it implies that there is a certain 
degree of degeneracy in the cosmological application of this gravitational modification: 
there are multiple combinations of $f_1$ and $f_2$ that give rise to the same function 
$g$.

Let us now define the variables
\begin{equation}\label{Var}
X=\frac{f_1-2 T  f_1'}{T(2 T  f_2'-f_2)},\quad \Omega=\frac{\rho _m}{6 \kappa H^2 },
\end{equation}
and the logarithmic time ${N}=\log a$. In this way Eqs. \rf{H0_2} tells us that 
$X=\Omega$, i.e. there is only one independent variable and the phase space is of 
dimension one. In the following we will choose  $\Omega$ as a key variable due to its 
straightforward physical meaning.

In the same way, Eq. \rf{H00}  can be written as
\begin{equation}\label{H00_1}
\begin{split}
& \dot{H} \left[\kappa  \left(f_1'+2T f_1''\right)-\rho _m \left(f_2'+2 T 
f_2''\right)\right]\\
&~~~~~~~~~+\frac{1}{4} \gamma \rho _m \left(f_2-2 T f_2'\right)=0,
\end{split}
\end{equation}
where we have used the relation $\dot{T}=-12 H \dot{H}$. In terms of the variables  
given in Eq. \rf{Var}, one may write Eq. (\ref{H00_1}) as
\begin{equation}
X [A(T)+\Omega  B(T)] \frac{ \dot{H}}{H^2}-\frac{3 \gamma}{2}\Omega=0,
\end{equation}
where $A(T)$ and $B(T)$ are defined as
\begin{equation}
\begin{split}
&A(T)=\frac{T\left(f_1'+2 T f_1''\right)}{f_1-2 T f_1'},\\
&B(T)=\frac{T^2\left(f_2'+2 T f_2''\right)}{f_1-2 T f_1'},
\end{split}
\end{equation}
respectively. Now, provided that the equation
\begin{equation}
\frac{f_1-2 T  f_1'}{T(2 T  f_2'-f_2)}=\Omega
\label{SApprox}
\end{equation}
can be inverted, the dynamics of the full cosmological system is described solely by a 
single equation for $\Omega$, namely
\begin{equation}\label{DynEqOm}
\DerN \Omega=-3 \gamma \Omega \left[1+\frac{1}{A(\Omega)+B(\Omega) \Omega}\right].
\end{equation}
The stability of the fixed points of this equation can be trivially obtained considering 
its second derivative with respect to $\Omega$, evaluated at the fixed points. The 
associated solution will be given integrating
\begin{equation}\label{RayGen}
\begin{split}
&\frac{ \dot{H}}{H^2}=-\frac{1}{p}\\
&\frac{1}{p}=-\frac{3 \gamma }{2 (A(\Omega_{*})+B(\Omega_{*}) \Omega_{*})},
 \end{split}
\end{equation}
where the subscript $*$ represents the value of $\Omega$ at the fixed point. Similarly to 
the usual method of the previous section, integrating and using the Bianchi identities we 
obtain
\begin{equation}
\begin{split}
& a=a_0(t-t_0)^p,\\
& \rho=\rho_0 (t-t_0)^{-3\gamma p}.
\label{powerlaw}
\end{split}
\end{equation}
Finally, from the structure of Eq. \rf{DynEqOm} we deduce that one of the possible solutions 
can be the vacuum ($\Omega=0$) one, but its existence depends on the form of the second 
factor. A second class of points, namely $\mathcal{B}$, can arise from the term in 
square parenthesis in \rf{DynEqOm}, representing a scaling solution with decelerated 
expansion. 

In the following three subsections, we apply the above general method in three specific 
examples based on different choices of $f_1$ and $f_2$. 

\subsection{Example 1 }\label{MSan2E1}

Let us consider the functions
\begin{equation}
f_1=-\frac{\alpha }{9}  T^5, \quad  f_2=-\frac{\beta }{3} T,
\end{equation}
where $\alpha,\beta$ are constants. In this case we have 
\begin{equation}
T=-\sqrt{\frac{ \beta \Omega }{\alpha}},
\end{equation}
and therefore Eq. \rf{DynEqOm} reads 
\begin{equation}
\DerN \Omega=2\gamma \Omega,
\end{equation}
which admits the fixed point $\Omega=0$. The solution associated to this fixed point can 
be found integrating \rf{RayGen} at the fixed point, once the functions $A$ and $B$ have 
been substituted. In particular, we find
\begin{equation}
\frac{\dot{H}}{H^2}=-\frac{1}{2}\gamma,
\end{equation}
which leads to the scaling solution
\begin{equation}
a=a_0 (t-t_0)^{\frac{2}{ \gamma}},
\end{equation}
that corresponds to accelerated expansion. It is easy to see that this point is 
an attractor for all parameter values. However, since this point 
corresponds to 
$T=0$, this 
form of the functions $f_1$ and $f_2$ cannot be accepted as physically motivated.

\subsection{Example 2}

As a second example, let us consider the functions
\begin{equation}\label{M2Ex3}
f_1= -\frac{\alpha}{6} T, \quad  f_2=-\frac{\delta \sqrt{-T}-\eta T}{2T},
\end{equation}
where $\alpha,\beta, \delta, \eta$ are positive constants. In this case we obtain
\begin{equation}
T=-\frac{36 \delta ^2 \Omega ^2}{(\alpha +6 \eta  \Omega )^2},
\end{equation}
and therefore Eq. \rf{DynEqOm} becomes
\begin{equation}\label{EqEx2-2}
\DerN \Omega=\frac{\gamma \Omega  (\alpha +6 \eta  \Omega )}{\alpha -6 \eta  \Omega },
\end{equation}
where without a significant loss of generality we have assumed that all constants are positive.  
This equation admits only one fixed points with non negative coordinates, namely 
$\Omega=0$ . The corresponding solutions are given by
\begin{align}
&a=a_0 (t-t_0)^{\frac{1}{3 \gamma}}
\end{align}
which represents a decelerated expansion scenario. This scaling solution is always unstable, and this implies the presence of asymptotic attractors. It is also interesting to note that since the \rf{EqEx2-2} posses a singularity in this case, like in the examples of the previous sections, there are initial conditions for which the system evolves towards a singularity. This implies that in general the models at hand can present dynamical instabilities.
\subsection{Example 3}

Finally, let us consider the functions
\begin{equation}
f_1= \log (-T)+2+\alpha,  \quad  f_2=-\frac{\beta  [3 \log (-T)+2]}{9 T},
\end{equation}
where $\alpha, \beta$ are constants. In this case we find  
\begin{equation}
T=-\exp\left(\frac{\alpha }{\beta  \Omega -1}\right),
\end{equation}
and therefore Eq. \rf{DynEqOm} reads as 
\begin{equation}
\DerN \Omega=-\frac{3 \gamma \Omega  (\beta  \Omega +1)^2}{\beta  \Omega  (\alpha +2+\beta 
 \Omega)
+1},
\end{equation}
which admits two non-negative fixed points, namely $\Omega=0$ and $\Omega=1/\beta$. The 
corresponding solutions are the de-Sitter expansion and the Friedmann solution 
(matter-dominated expansion). The first one is stable for any values of the parameters 
$\alpha, \beta$ and positive $w$. However, the Friedmann point is not hyperbolic: it is an 
attractor for $\Omega>1/\beta$ and a repeller for $\Omega<1/\beta$. This point 
will give a consistent value of the parameter $\Omega$ for $\beta>1$. Hence, for this 
specific model the transition between matter-dominated epoch and de-Sitter dark-energy 
era can be described naturally.

Like in the previous example, also in this case the phase space presents singular points.
Some initial condition lead to these singularities showing also in this case the possibility of instabilities of this class of theories.

\section{Discussions and final Remarks}\label{Conclusions}

Recently, a generalized $f(T)$ gravity theory with a nonminimal coupling between matter 
and the torsion scalar was proposed in \cite{NFT}. From the physical point of view, in this 
theory, matter is not just a passive component in the space-time continuum; rather, it 
plays an active role in the overall gravitational dynamics, which is strongly modified due 
to the supplementary interaction between matter and geometry 
\cite{NFT,Bertolami:2007gv,Harko:2008qz}. Moreover, the major advantage of the $f(T)$-type 
models, namely that the field equations are of second order, is not modified by the 
torsion-matter coupling  \cite{Cai:2015emx}.

In this work we have performed a detailed phase-space analysis for this scenario, 
proposing two new approaches. The first one, although in principle more general, turns out 
to lead to complications, not only in terms of the dynamical equations, but also 
in terms of external constraints that have to be considered during the dynamical 
analysis. Hence, we have developed a second method, which can be alternatively used in 
cases where the above complexities make the application of the first method difficult, 
since in this approach all constraints are included in the initial equations. Definitely, 
the physical results do not depend on the specific mathematical method used, and thus one 
can choose the easier method according to the specific problem.

We have analyzed six specific models of non-minimally matter-torsion coupled cosmology. 
In general we have extracted fixed points corresponding either to dark-matter dominated, 
scaling decelerated solutions, or to dark-energy dominated accelerated solutions. 
Additionally, note that there is a small parameter region in which the model can 
experience the transition from the matter epoch to dark-energy era. This feature is in 
agreement with the observed universe evolution, and cannot be easily obtained in dark 
energy models. It is also worth mentioning that in both the approaches reveal  that there 
are sets of initial conditions  for which the models evolve towards singular states for the 
equations. This tendency is a signal of  instability related with certain specific forms of 
the coupling functions and this feature constraints even more the viable parameter 
space of these models. 

We mention that in the present analysis we have focused on the finite 
regions of the phase space, without examining the possible non-trivial behavior at 
infinity, since such an extended investigation would lie beyond the scopes of this work, 
which is just to reveal the capabilities of the theory in producing a 
cosmology in agreement with observations. As we showed this is possible, and hence  
non-minimally matter-torsion coupled gravity may be a successful candidate for the 
description of Nature.

\begin{acknowledgments}
SC was supported by  the Funda\c{c}\~{a}o para a Ci\^{e}ncia e Tecnologia through project 
IF/00250/2013. FSNL acknowledges financial support of the Funda\c{c}\~{a}o para a 
Ci\^{e}ncia e Tecnologia through an Investigador FCT Research contract, with reference 
IF/00859/2012, and the grants EXPL/FIS-AST/1608/2013 and PEst-OE/FIS/UI2751/2014.
GO wishes to thank the Laboratory for High-Energy Physics (LAFEX) of the Brazilian Centre 
for Research in Physics, CBPF, for hospitality, PCI/MCT-CNPq and CAPES (PNPD) for 
financial support.
\end{acknowledgments}

\end{document}